# Canonical Demon Monte Carlo Renormalization Group

M. Hasenbusch[a], K. Pinn, and C. Wieczerkowski[*b]

[a]DAMTP, Silver Street, Cambridge, CB3 9EW, England

[b]Institut für Theoretische Physik I, Wilhelm-Klemm-Str. 9, D-48149 Münster, Germany

We describe a method to compute renormalized coupling constants in a Monte Carlo renormalization group calculation. It can be used, e.g., for lattice spin or gauge models. The basic idea is to simulate a joint system of block spins and canonical demons. Unlike the Microcanonical Renormalization Group of Creutz et al. it avoids systematical errors in small volumes. We present numerical results for the $O(3)$ nonlinear $\sigma$-model.

In the Monte Carlo Renormalization Group (MCRG) [1] block spin transformations (BST) are analyzed by Monte Carlo (MC) simulations. In its original formulation this is done in terms of blocked observables suitable for the computation of critical properties. This includes the determination of critical exponents from linearized renormalization group (RG) transformations [2] and the matching method for the calculation of the $\Delta\beta$ function [3]. Another formulation aims at the computation of effective actions generated by a BST. Here the difficulty is that effective actions can be complicated with many couplings. In addition they can be hard to compute.

Creutz et al. propose to compute renormalized couplings in a microcanonical demon simulation [4]. Their method rests on the equivalence of a microcanonical and a canonical ensemble in the thermodynamic limit. In small volumes it has systematic errors. This disadvantage can be overcome by making the demons canonical [7]. The resulting method requires also just a standard update program for its implementation.

## 1. Description of the Method

We consider a lattice spin system with spins $\phi_x$, action $S$, and Boltzmann factor is $\exp(-S)$. The action is assumed to have the form $S = -\sum_\alpha \beta_\alpha S_\alpha$ with interactions $S_\alpha$ and couplings $\beta_\alpha$. We then introduce an auxiliary demon system with action $S_D = \sum_\alpha \beta_\alpha d_\alpha$ and identical

*Speaker at the conference

couplings $\beta_\alpha$. The demons $d_\alpha$ are real numbers in the interval $[0, d_{max}]$. Consider the joint partition function

$$Z = \left(\prod_\alpha \int_0^{d_{max}} d\, d_\alpha\right) \int D\phi \, \exp(-S - S_D) \quad (1)$$

of spins and demons. It factorizes. The expectation value of the demon variables $d_\alpha$ is

$$<d_\alpha> = \frac{1}{\beta_\alpha}\left(1 - \frac{\beta_\alpha d_{max}}{\exp(\beta_\alpha d_{max}) - 1}\right). \quad (2)$$

Once it is known we can compute $\beta_\alpha$.

Let us turn to a numerical RG transformation. We can simulate the effective theory without explicit knowledge of the effective action. We simulate a spin model with a given action, and apply a certain blocking rule to the configurations on the fine grid to produce block spin configurations on a coarser grid. The blocked configurations are distributed according to an effective Boltzmann weight. The effective action can then be computed by a simulation of the above joint partition function.

This joint simulation can be realized as follows. Assume that the effective action is of the above general form. First perform microcanonical updates of the joint system. These updates do not change the differences $S_\alpha - d_\alpha$. Since also the Boltzmann weight is unchanged, knowledge of the $\beta_\alpha$ is not required for the update. Then replace the block spin configuration by a statistically independent one. The second step is ergodic and fullfils detailed balance. Statistical independence of the block configurations can be assured if the configurations are sufficiently well separated in computer time.

## 2. Numerical Results

We implemented this method for the $O(3)$ invariant vector model in two dimensions. We took an action $S = -\sum_{\alpha=1}^{12} \beta_\alpha \sum_x S_{x,\alpha}$ with 12 interaction terms $S_{x,\alpha} = \frac{1}{2} \sum_{y \in Y_{x,\alpha}} (s_x \cdot s_y)^n$. The $s_x$ are 3-vectors of unit length. $Y_{x,\alpha}$ consist of all lattice points obtained by the obvious symmetry operations from a representative lattice vector $v$, and $n$ takes values in $\{1, 2, 3\}$. Table 1 gives $n$ and $v$ for $\alpha = 1...12$. As a first test we reproduced given couplings. The model was simulated with some ad hoc chosen set of couplings on an $8^2$ and a $16^2$ lattice. The updates were performed with a tunable overrelaxation algorithm. The demon-spin update was done as follows. First, a new value for a single spin was proposed. Then we checked whether $d_\alpha$ plus the change of $S_\alpha$ was within $[0, d_{max}]$. If this was the case the system was updated.

Table 1
Labelling of the 12 interaction terms $S_{x,\alpha}$

| $\alpha$ | 1   | 2   | 3   | 4   | 5   | 6   |
|----------|-----|-----|-----|-----|-----|-----|
| $v$      | 1,0 | 1,1 | 2,0 | 2,1 | 3,0 | 1,0 |
| $n$      | 1   | 1   | 1   | 1   | 1   | 2   |
| $\alpha$ | 7   | 8   | 9   | 10  | 11  | 12  |
| $v$      | 1,1 | 2,0 | 2,1 | 3,0 | 1,0 | 1,1 |
| $n$      | 2   | 2   | 2   | 2   | 3   | 3   |

We employed 100 independent demon systems in order not to waste too many block spin configurations. We generated a sequence of spin configurations, two successive configurations being separated by a number of sweeps large compared to the autocorrelation time divided by 100. The configurations for the 100 demon systems were then successively taken from this sequence. After a replacement of the spin configuration, we always performed one lattice spin-demon updating sweep. After a full cycle through the demon systems the measured demon values were averaged over the 100 systems and written to disk.

Table 2
Reproduction of the original coupling constants

|            | couplings | $L = 8$       | $L = 16$      |
|------------|-----------|---------------|---------------|
| $\beta_1$  | 1.30      | 1.3010(23)    | 1.2999(12)    |
| $\beta_2$  | 0.35      | 0.3481(12)    | 0.3485(8)     |
| $\beta_3$  | 0.01      | 0.0105(9)     | 0.0098(6)     |
| $\beta_4$  | 0.02      | 0.0190(7)     | 0.0192(4)     |
| $\beta_5$  | 0.004     | 0.0045(6)     | 0.0039(5)     |
| $\beta_6$  | −0.200    | −0.2045(30)   | −0.2002(18)   |
| $\beta_7$  | −0.080    | −0.0808(21)   | −0.0808(12)   |
| $\beta_8$  | −0.020    | −0.0201(11)   | −0.0188(6)    |
| $\beta_9$  | −0.01     | −0.0086(6)    | −0.0085(4)    |
| $\beta_{10}$ | −0.005  | −0.0045(8)    | −0.0038(6)    |
| $\beta_{11}$ | 0.02    | 0.0248(26)    | 0.0220(14)    |
| $\beta_{12}$ | 0.01    | 0.0131(19)    | 0.0114(12)    |

Table 3
Truncation experiment

|           | couplings | $L = 8$     | $L = 16$    |
|-----------|-----------|-------------|-------------|
| $\beta_1$ | 1.30      | 1.1399(10)  | 1.1408(7)   |
| $\beta_2$ | 0.35      | 0.2993(6)   | 0.3014(4)   |
| $\beta_3$ | 0.01      | − 0.0020(6) | −0.0017(4)  |
| $\beta_4$ | 0.02      | 0.0138(3)   | 0.0149(2)   |
| $\beta_5$ | 0.004     | 0.0019(4)   | 0.0022(2)   |

Our results are presented in Table 2. In addition to the reproduction of the 12 original couplings we studied the truncation to a subset of 5 couplings (Table 3). For the $L = 8$ lattice we performed 20000 updates of the 100 demon systems. In the $L = 16$ case we made 10000 updates. The results show the couplings are reproduced within error bars.

For the original set of couplings we measured a correlation length $\xi = 57.8(4)$. The correlation length of the truncated coupling set ($L = 16$) turned out to be smaller than the original one. We found $\xi_{trunc} = 36.7(2)$. Note, however, that a naive truncation of the original set of couplings to the first five couplings leads to $\xi = 195.5 \pm 2.3$ on an $L = 400$ squared lattice.

Next we studied renormalization group transformations, starting from the standard action with nearest neighbour coupling only. We used a blocking rule that we call "dressed decimation".

Table 4
Results for the first four RG steps, starting from the standard action with $\beta_1 = 1.9$. The last line gives the correlation lengths for the effective theories. The columns give the results of the subsequent RG steps

| RG step | 0 | 1 | 2 | 3 | 4 |
|---|---|---|---|---|---|
| $\beta_1$ | 1.9 | 1.4619(15) | 1.1764(11) | 0.9785(16) | 0.8201(14) |
| $\beta_2$ | 0.0 | 0.2691(9) | 0.3117(8) | 0.2892(8) | 0.2528(8) |
| $\beta_3$ | 0.0 | $-0.0123(6)$ | 0.0107(5) | 0.0226(6) | 0.0272(4) |
| $\beta_4$ | 0.0 | 0.0074(4) | 0.0132(3) | 0.0182(3) | 0.0195(4) |
| $\beta_5$ | 0.0 | 0.0043(5) | 0.0031(4) | 0.0032(5) | 0.0034(4) |
| $\beta_6$ | 0.0 | $-0.2213(19)$ | $-0.2611(13)$ | $-0.2257(14)$ | $-0.1689(12)$ |
| $\beta_7$ | 0.0 | $-0.0800(12)$ | $-0.1027(10)$ | $-0.0985(13)$ | $-0.0766(9)$ |
| $\beta_8$ | 0.0 | 0.0003(9) | $-0.0057(6)$ | $-0.0104(7)$ | $-0.0107(8)$ |
| $\beta_9$ | 0.0 | $-0.0022(5)$ | $-0.0011(5)$ | $-0.0077(5)$ | $-0.0081(5)$ |
| $\beta_{10}$ | 0.0 | $-0.0005(6)$ | $-0.0011(5)$ | $-0.0014(7)$ | $-0.0015(7)$ |
| $\beta_{11}$ | 0.0 | 0.0659(17) | 0.0898(13) | 0.0748(18) | 0.0470(15) |
| $\beta_{12}$ | 0.0 | 0.0327(13) | 0.0440(11) | 0.0425(15) | 0.0301(12) |
| $\xi$ | 121.2(6) | 57.0(4) | 24.5(2) | 10.47(5) | 4.78(2) |

All lattice points that have coordinates $x = (i, j)$ with $i$ and $j$ even are identified with block sites. The block spin $s'$ at site $x$ is then defined as

$$s'_x = \frac{ws_x + \frac{1}{4}(1-w)\sum_{y.nn.x} s_y}{|ws_x + \frac{1}{4}(1-w)\sum_{y.nn.x} s_y|}, \quad (3)$$

where the sum is over the nearest neighbours of $x$. Tests with the massless free field theory revealed that the choice $w = 0.8$ is a good one in the sense that the effective actions and especially the fixed point action had good locality properties.

As a first test we started with $\beta_1 = 1.9$, and all the other couplings put to zero. The correlation length for this action is 121.2(6) [5]. We then blocked a $32^2$ lattice down to a $16^2$ lattice. The effective 12 couplings were then determined from the demon expectation values and used as input for the next iteration. Our results for the first 4 steps are presented in Table 4. The last line shows correlation length estimates for the effective theories. The correlation length should theoretically be exactly halved each step. The significant deviations indicate that the action is not sufficiently well approximated by our ansatz. Note however that there is a very good decay of the couplings for given $n$ with increasing distance of the spins. We conclude that local interactions with higher $n$ and also interactions with more than two spins cannot be left out. (E.g., 4-point operators proved to be important in the study of Hasenfratz et al. [6] using the classical approximation.)